\documentclass[prb,twocolumn,showpacs]{revtex4}

\usepackage{graphicx}
\usepackage{dcolumn}
\usepackage{amsmath}
\usepackage{times}
\newcommand{\LS}{La$_{2/3}$Sr$_{1/3}$MnO$_{3}$ }
\newcommand{\LC}{La$_{2/3}$Ca$_{1/3}$MnO$_{3}$ }

\begin{document}
\title{Cooperative dynamics in doped manganite films: \\
phonon anomalies in the ferromagnetic state}

\author{Ch.~Hartinger}
\affiliation{EP V, Center for Electronic Correlations and
Magnetism, University of Augsburg, 86135 Augsburg, Germany}

\author{F.~Mayr}
\affiliation{EP V, Center for Electronic Correlations and
Magnetism, University of Augsburg, 86135 Augsburg, Germany}

\author{A.~Loidl}
\affiliation{EP V, Center for Electronic Correlations and
Magnetism, University of Augsburg, 86135 Augsburg, Germany}

\author{T.~Kopp}
\affiliation{EP VI, Center for Electronic Correlations and
Magnetism, University of Augsburg, 86135 Augsburg, Germany}

\date{\today}

\begin{abstract}
We present optical measurements of phononic excitations in \LC
(LCMO) and \LS (LSMO) thin films covering the full temperature
range from the metallic ferromagnetic to the insulating
paramagnetic phase. All eight phonons expected for the R$\bar{3}$c
symmetry in LSMO and 17 out of the expected 25 phonons for the
Pnma symmetry in LCMO have been determined. Close to the
ferromagnetic-to-paramagnetic transition both compounds reveal an
anomalous behavior but with different characteristics. Anomalies
in the phononic spectra are a manifestation of the coupling of
lattice degrees of freedom (DOF) to electronic DOF. Specifically,
the low-frequency external group proves to be an indicator for
lattice modifications induced by electronic correlations. The
enhanced electron-phonon coupling in LCMO is responsible for
Fano-like interference effects of distinct phonon modes with
electronic continuum excitations: we observe asymmetric phonon
line shapes, mode splitting and spectral weight transfer between
modes.
\end{abstract}


\maketitle
\section{Introduction}
With the enormous progress of research on colossal
magnetoresistance (CMR) manganites\cite{Helmolt93} over the last
decade it has been realized that besides charge and spin also {\it
lattice degrees of freedom}~\cite{Millis95} play a significant
role in the formation of the CMR and the metal-insulator
transition (MIT). Special attention focused on the optimal doped
manganites, characterized by the largest CMR effects---in
particular, La$_{2/3}$Ca$_{1/3}$MnO$_{3}$ (LCMO). As the
temperature range in which the CMR effect and the MIT are observed
is close to room temperature, these compounds have become
potential candidates for technical applications.

The phonon excitation spectrum has been studied in detail for
undoped LaMnO$_{3}$~\cite{Fedorov99,Smirnova99} which is, below
approximately 750~K, in a Jahn-Teller (JT) distorted orthorhombic
(Pnma) structure. A theoretical analysis of the complete set of
phonon modes in the doped compounds is still missing. LCMO has an
orthorhombic  symmetry whereas the Sr-doped compound,
La$_{2/3}$Sr$_{1/3}$MnO$_{3}$ (LSMO), is rhombohedrally distorted.
The infrared-active phonons of undoped rhombohedral (R$\bar{3}$c)
LaMnO$_{3}$ were identified~\cite{Abrashev99} however these
zero-doping investigations cannot make up for experiments at
finite doping. In fact, the local Mn-O geometry depends strongly
on the size of the ions and doping
concentration~\cite{Hwang95,Goodenough63,Urushibara95} which
accounts for the structural differences between LCMO and LSMO.

Experimental studies of two single crystals, undoped and with
$\sim 8$\% Sr-doping, were performed by Paolone {\it et
al.}\cite{Paolone00}  They discussed the assignment of all
infrared active phonon modes in LaMnO$_{3}$ at low temperature.
However, to the best of our knowledge, there are no reports on
phonon modes in the metallic phase of LSMO. For bulk samples of
LCMO, three sets of broad phonon bands are observed which reveal
external-, bending- and stretching-type lattice vibrations, as
classified in the cubic symmetry.~\cite{Kim96,Boris99}

Apart from providing a fingerprint of the lattice structure,
phononic spectra can indicate the relevance of electronic
correlations. The dominant effect of mobile electrons on the
lattice vibrations is generically a suppression of phononic
resonances due to screening. For metals, this may render the
phonons unobservable in far-infrared (FIR) optical spectroscopy.
However, since the polycrystalline samples are poor metals, and
even more so the investigated thin films, phonon resonances are
still clearly resolved even in the conducting low-temperature
ferromagnetic phase.

In LCMO a significant shift of the stretching mode near the MIT
was found~\cite{Kim96} and associated with a reduced screening
through charge carriers close to the MIT. Although the observed
shifts of phonons in LCMO are evidently related to electronic
correlations since they come along with the MIT it is not obvious
that they are explained exclusively by a reduced screening. In
particular, not all phonon modes display the shift and,
furthermore, in LSMO we observe phononic shifts which are tied to
the ferromagnetic-to-paramagnetic (FM-PM) transition, not the MIT
(see Sec.~\ref{sec:comp}). In LCMO the FM-PM transition and the
MIT coincide whereas in LSMO they are well separated. A
comparative study of LCMO and LSMO should address and illuminate
the fundamental question if the phononic excitations couple to
spin degrees of freedom.

Since the phonon resonances are well pronounced in the optical
spectroscopy measurements of the metallic manganite films they
provide a unique opportunity to analyze the consequences of a
coupling of lattice degrees of freedom (DOF) with spin and charge.
Correlation of lattice DOF with charge DOF has to be in an weak
(LSMO) to strong coupling range (LCMO) as, for example, the
polaronic excitations in the mid-infrared (MIR) evidence a
coupling in this range.\cite{Hartinger04a} Also neutron
diffraction measurements for LCMO revealed an anomalous volume
thermal expansion at the paramagnetic-to-ferromagnetic
MIT,\cite{Radaelli95,Teresa96,Dai96} which has to originate from a
sufficiently strong coupling. We will elaborate further on the
consequences of a strong phonon-electron coupling for the phononic
resonances in this paper and we will identify the characteristics
in the FIR spectra signifying cooperative dynamics not only of
lattice and electronic charge but also of lattice and spin degrees
of freedom.

The paper is organized as follows: Sec.~\ref{sec:ExpDetails} deals
with the experimental details, sample characterization and data
analysis. In Sec.~\ref{sec:TempR}, we present reflectivity
measurements of the infrared active phonons in thin films of LSMO
and LCMO at temperatures ranging from the insulating paramagnetic
(PM) to the metallic ferromagnetic (FM) phase. In Sec.~\ref{sec:HT}
we give a comprehensive analysis of phonons from the respective
optical data. The spectra confirm the anticipated mode structure
for the respective symmetry of the lattice and prove the high
quality of the thin films. In Sec.~\ref{sec:comp} we elaborate on
the  close relation between lattice and spin degrees of freedom by
studying the temperature dependent shift of phonon modes across
the ferromagnetic transition.
In Sec.~\ref{sec:Fano}, we discuss Fano-like modifications of the
line shape of distinct phonon modes in LCMO which testify a
significant coupling of the lattice excitations to the electronic
continuum in the Ca-doped manganite films. The interference of
nearly degenerate modes with electronic excitations may result in
spectral weight transfer and level repulsion which we discuss
within a phenomenological approach for coupled phononic and
electronic modes.
A summary of our findings and conclusions are presented in
the last section, Sec.~\ref{sec:conclusion}.

\section{Experimental Details and Sample Characterization}
\label{sec:ExpDetails}

The films for this investigation have been prepared using a
standard pulsed laser deposition technique.\cite{Christey94}
LCMO was grown to a thickness of  200~nm  onto  NdGaO$_{3}$ single
crystalline substrates. LSMO~\#1 with a thickness of $d=300$~nm
and LSMO~\#2 with $d=400$~nm  were deposited on
(LaAlO$_{3}$)$_{0.3}$(Sr$_{2}$AlTaO$_{5}$)$_{0.7}$.
X-ray diffraction of LCMO revealed an orthorhombic structure with a preferred
growth along the [110] axis while LSMO was confirmed to be in a rhombohedral
structure with preferred growth along the [100] axis.

The infrared reflectivities of the sample and the pure substrate
were measured using a combination of Fourier transform
spectrometers, Bruker IFS 113v and IFS 66 v/S, to cover the
frequency range from 50 to 40000~cm$^{-1}$. Temperature dependent
measurements from 6~to 295~K were performed with a He-cryostat.
Higher temperatures were measured in a home made oven in which the
sample was exposed to a continuous flow of heated nitrogen gas. To
obtain the optical conductivity $\sigma$ a Kramers-Kronig (KK)
analysis was performed which included the results from
submillimeter spectroscopy between 5 and 30~cm$^{-1}$. Applying
the Fresnel optical formulas for the complex reflectance
coefficient of the substrate-film system, the optical conductivity
for the films was calculated.\cite{Heavens91}

Magnetization measurements were carried out between 2 and 400~K by
using a commercial Quantum Design SQUID magnetometer. The dc
resistance measurements were performed by a standard four-probe
method. The temperature dependence of the ac resistancee was done
in a Mach-Zehnder interferometer arrangement, which allows
the measurements of both, transmission and phase shift. Applying the
Fresnel formulas, the ac resistance was determined directly
without any approximations.

\begin{figure}[b]
\vspace{4mm} \centering
\includegraphics[width=.47\textwidth,clip,angle=0]{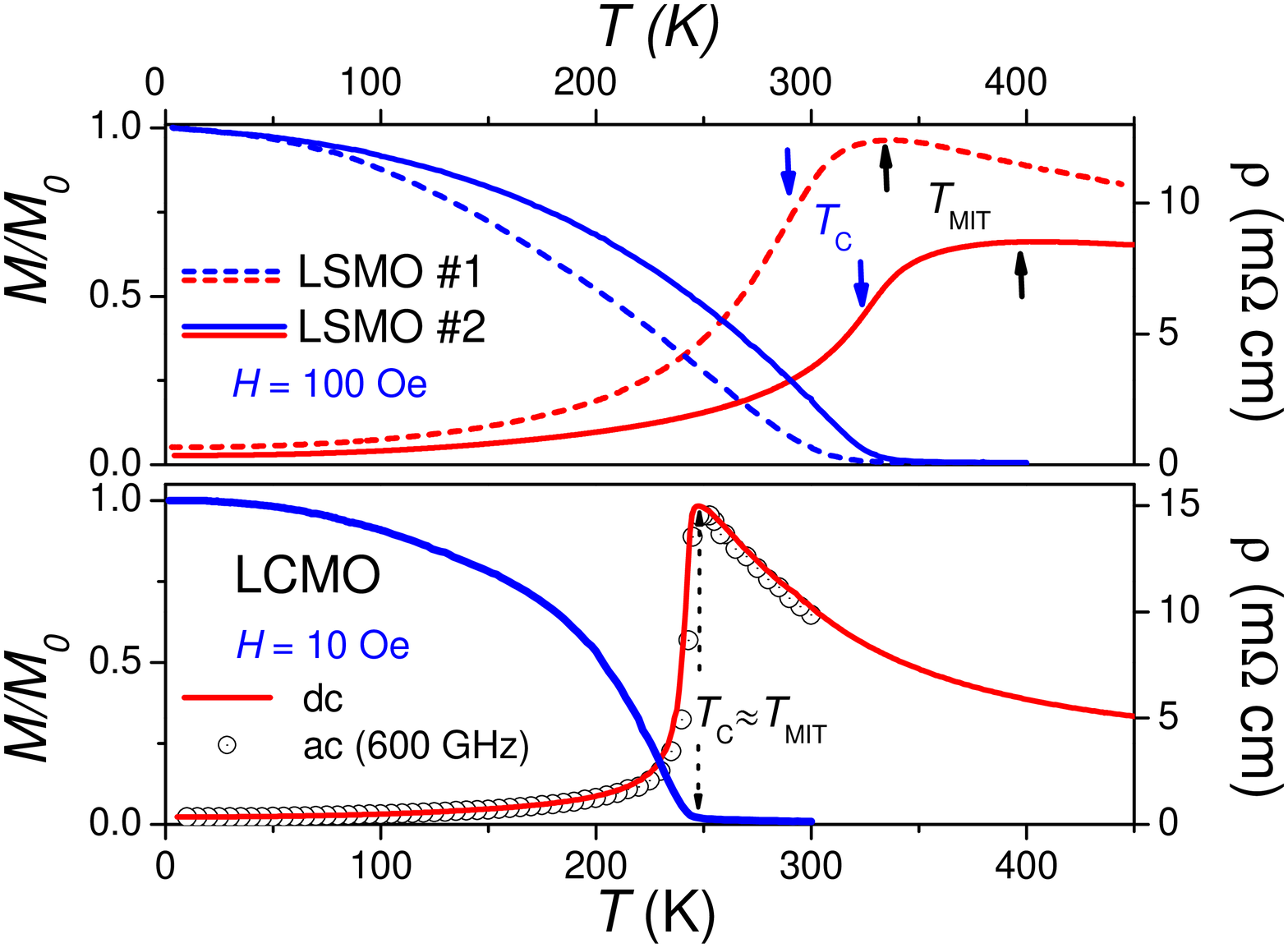}
\caption[]{\label{Ca-squid-DC-AC}Upper panel (LSMO): Temperature
dependence of the magnetization at $H=100$~Oe (left scale) and
DC-resistivity (right scale). Lower panel (LCMO): Magnetization
at $H=10$~Oe (left scale); DC- and AC-resistivity at $\nu=600$~GHz
(right scale).}
\end{figure}

Frequently, manganite films exhibit FM transition temperatures
$T_{C}$ lower than those reported for the bulk materials with the
same nominal composition. The differences can be attributed to
grain boundaries, strain, microstructure depending on the details
of the growth process and on the lattice parameters of the
substrate.\cite{Worledge96,Gommert99} They all result in a
variation of the strength of the lattice distortion, which
determines the structure and, accordingly, the electrical and
magnetic behavior. It is well known, that differences between bulk
samples and thin films can occur, depending on the quality of the
films. However, we observe that the overall optical characteristics of the films
are in good agreement with bulk properties.

\begin{table}[t]

\caption[]{Metal-insulator ($T_{\rm MIT}$), magnetic
($T_{C}^{{}^{\chi}}$) and electric ($T_{C}^{{}^{\rho}}$)
transition temperatures of the LCMO and LSMO films.} \label{tabfilmtc}

\vspace{.20cm} \centering
\begin{tabular}{llcccccllll } \hline\hline
  &&  $T_{\rm MIT}$ (K)&&  $T_{C}^{{}^{\rho}}$ (K)&&
$T_{C}^{{}^{\chi}}$ (K) && sample    \\ \hline
  LCMO ($d$=200~nm)  &&245&&  243 && 242 && LCMO  \\
  LSMO ($d$=300~nm) &&338&&  287 && 300 && LSMO~\#1 \\
  LSMO ($d$=400~nm) &&401&&  328 && 345 && LSMO~\#2 \\ \hline\hline
\end{tabular}
\end{table}

To probe the electrical and magnetic properties of the different
films we determined the characteristic transition temperatures. We
define $T_{C}^{{}^{\chi}}$ as the magnetic Curie temperature,
$T_{C}^{{}^{\rho}}$ as the inflection point\cite{Fisher68} and
$T_{\rm MIT}$ is taken from the maximum in the dc and ac
resistivity curve. We summarize the results of the transition
temperatures for both compounds in Tab.~\ref{tabfilmtc}. Since
$T_{C}^{{}^{\chi}}$ and $T_{C}^{{}^{\rho}}$ are relatively
similar, we will use the notation $T_C$ for either one if not
stated otherwise. To identify the MIT and FM-PM transition of the
thin film samples, we display magnetization and dc as well as ac
resistivity in Fig.~\ref{Ca-squid-DC-AC}. As expected, $T_C$
nearly coincides with $T_{\rm MIT}$ for LCMO.

\section{Experimental Results and Discussion}
\subsection{Temperature dependence of the reflectivity}
\label{sec:TempR}

\begin{figure}[b]
\centering
\includegraphics[width=.5\textwidth,clip,angle=0]{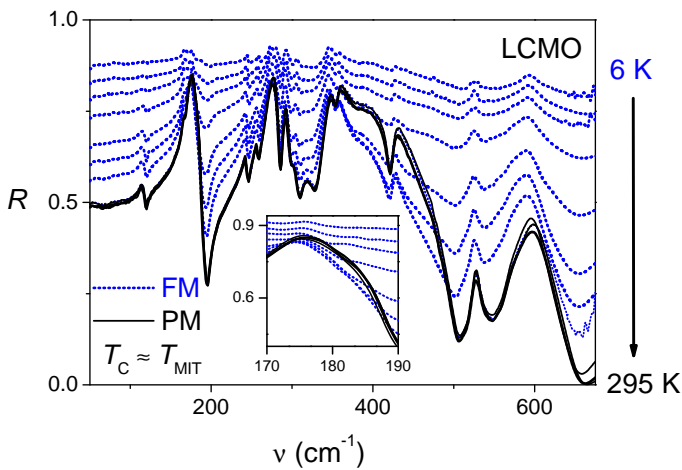}
\vspace{0mm} \caption[]{\label{Ca-r-temp} Temperature dependence
of the reflectivity $R$ of LCMO. The dotted lines represent
temperatures in the FM phase, the solid lines in the PM phase.
Above the crossover the intensity of $R$ is approximately
temperature-independent. The anomalies of some phonons at $T_{C}$
evidence the significant role of electronic DOF in the
modification of the lattice and its vibrations at the FM-PM MIT.
Most pronounced is the enhancement of the reflectivity near the
bending modes but an increment of $R$ is also observable for the
higher-frequency external modes, cf.~inset.}
\end{figure}

The frequency-dependent reflectivity of LCMO is presented in
Fig.~\ref{Ca-r-temp} for temperatures in the metallic  FM (dotted
lines) and in the insulating PM (solid lines) phase. Above the MIT
(245~K) the shape and intensity of $R$ is temperature independent.
With the temperature increasing from 6~K\ to 295~K, the weight of
several phonon modes gradually increases and a typical insulating
behavior with an energy gap of about 680~cm$^{-1}$ develops.
Variations in position and intensity of some characteristic phonon
modes are apparent.

\begin{figure}[t]
\centering
\includegraphics[width=.5\textwidth,clip,angle=0]{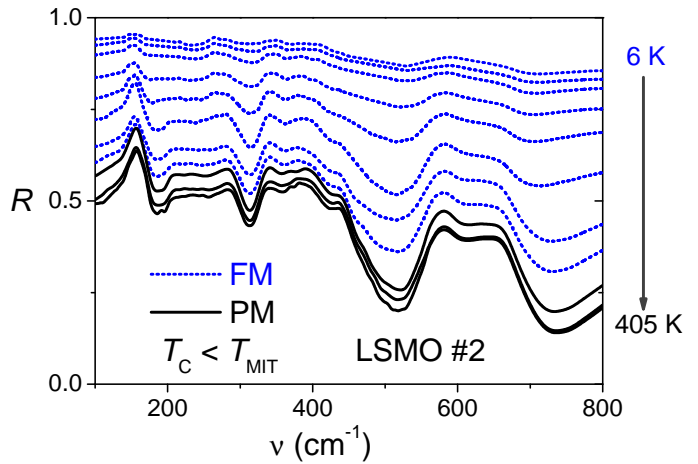}
\vspace{0mm} \caption[]{\label{Sr-r-temp}Temperature dependence of
the reflectivity $R$ of LSMO~\#2. The dotted lines represent
temperatures in the FM phase, the solid lines in the PM phase.
With increasing temperature $R$ gradually decreases up to $T_{\rm
MIT}=401$~K, which is above $T_{C}=345$~K.}
\end{figure}

The overall distinction between the reflectivity spectra of LCMO
and LSMO is striking (compare Figs.~\ref{Ca-r-temp} and
\ref{Sr-r-temp}). The less complex structure of the phonon
spectrum of LSMO, see Fig.~\ref{Sr-r-temp}, is evidence for the
higher symmetry of the unit cell. Above 400~K the spectra hardly
change. The reflectivity in LSMO \#1 is slightly lower
than in LSMO \#2, otherwise the spectra are essentially identical.

In addition, the LCMO spectra display phonon anomalies at $T_{C}$,
a feature which is entirely missing in LSMO. X-ray and neutron
powder diffraction studies demonstrate an anomalous volume thermal
expansion for the Ca-doped compound at the
MIT.\cite{Radaelli95,Teresa96,Dai96} Dai \textit{et
al.}~\cite{Dai96} show that this anomaly originates from
variations of the Mn and O positions. The modified Mn-O bond
lengths and bond angles are expected to cause anomalies of those
phonon modes which are particularly sensitive to variations of the
Mn-O geometry. Contrary to LCMO, LSMO exhibits no anomalies in the
lattice parameters and in the unit cell volume around
$T_{C}$.\cite{Martin96,Mellergard00}

\subsection{Phonon modes in LCMO and LSMO}
\label{sec:HT}

The different lattice structure of LSMO and LCMO has an impact on
the optical conductivity, and on the number of observed phonon
modes. Approximately, the phonon spectra of manganites can be
separated into external ($\sim$\,185 cm$^{-1}$), bending
($\sim$\,350 cm$^{-1}$) and stretching ($\sim$\,550 cm$^{-1}$)
modes with respect to cubic (Pm3m) symmetry. Depending on ion size
and doping concentration, these triply degenerate modes split into
pairs of non-degenerate ($A$) and doubly degenerate ($E$) modes,
and, moreover, they become broader and
overlap.\cite{Paolone00,Mayr03} Furthermore, due to the larger
unit cell additional modes emerge.

\begin{figure}[t]
\vspace{0mm} \centerline{ \ \ \ \ \ \
\includegraphics[width=.5\textwidth,clip,angle=0]{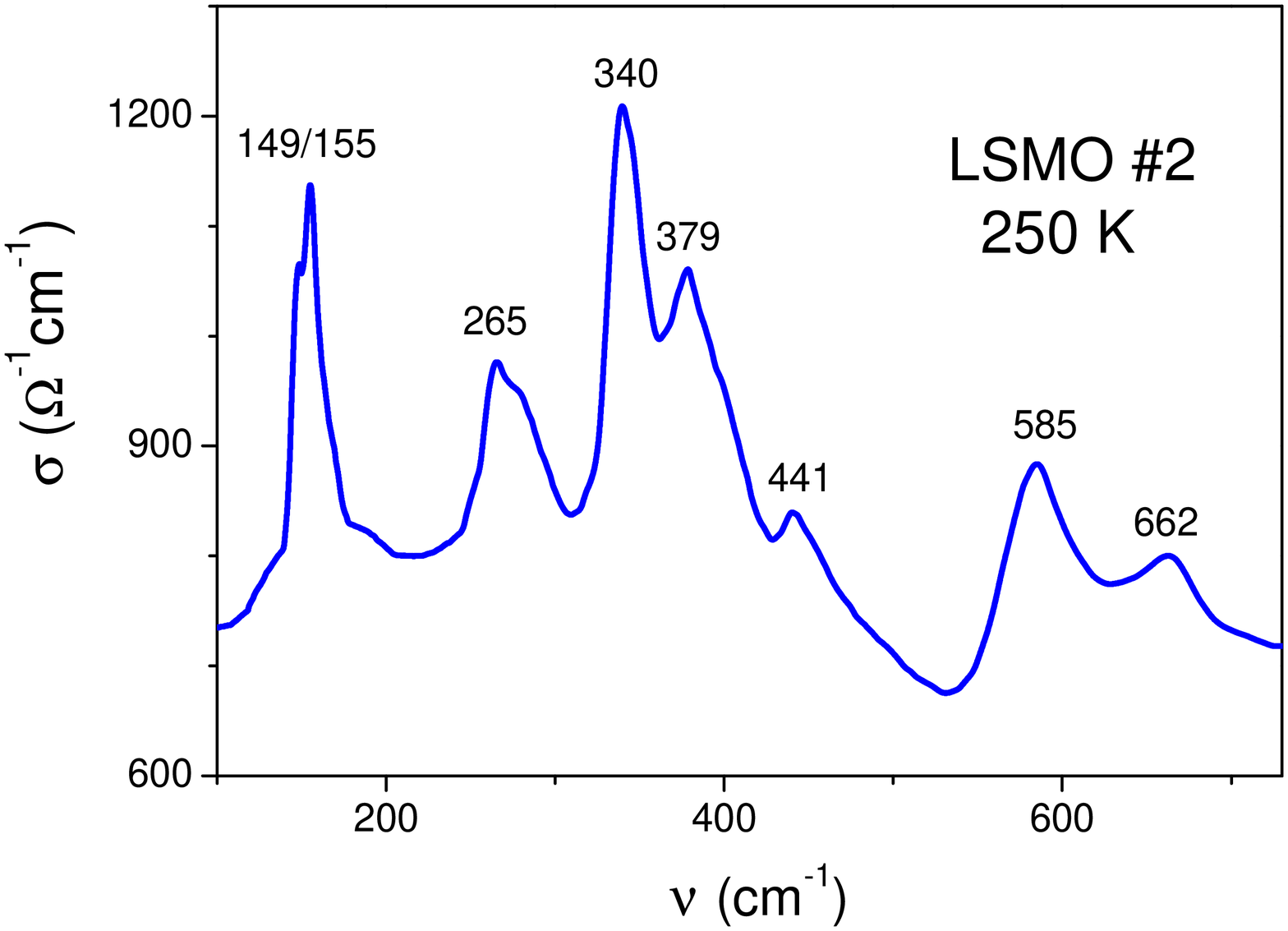}}
\caption[]{\label{Sr-Phononen} Optical conductivity of the
LSMO~\#2 film at 250~K.  Those phonon modes which are expected in
R$\bar{3}$c symmetry are labelled by numbers. They correspond to
the phonon frequencies in units of cm$^{-1}$.}\end{figure}

LSMO has a rhombohedral symmetry with space group R$\bar{3}$c. The
tilt of the octahedra results in a doubling of the cubic unit
cell, and 8 ($5 E_{u}$ and $3 A_{2u}$) modes are IR active. LCMO
has the Pnma structure with an orthorhombic unit cell containing
four cubic units. Hence, 25 modes should be visible in the FIR
spectra. In this section, we focus on spectra at high temperatures
where the phonon peaks are better resolved.

\begin{table}[b]
\caption[]{Correspondence between calculated \cite{Abrashev99} and
measured phonon modes for the space group R$\bar{3}$c. The
measured frequencies (in cm$^{-1}$) refer to  LSMO~\#2 at
$T=250$~K, and the values in parenthesis to neutron
scattering~\cite{Reichardt99} with aLa$_{0.7}$Sr$_{0.3}$MnO$_{3}$
sample. \label{tabmoden}}
\begin{center}
\begin{tabular}{clccccllcllc} \hline \hline
 &\multicolumn{3}{c}{calculated}      &&    &  &\multicolumn{3}{c}{measured}     &\\
 &$A_{2u}$ &  & $E_{u}$  &&    assignment &  &$A_{2u}$ &  & $E_{u}$  &\\ \hline
 &         &  & 317      && vibration (Mn)& &         &  & 379      & \\
 &162      &  & 180      && external& &149      &  & 155      & \\
 &310      &  & 357      && bending&  &340  (336)    &  & 441  (424)    &\\
 &641      &  & 642      && stretching&  &585 (576)     &  & 662      &\\
 &         &  & 240      && torsional&  &         &  & 265      &\\  \hline \hline
\end{tabular}
\end{center}
\end{table}

The optical conductivity, $\sigma$, of LSMO~\#2 at 250~K is
obtained from the KK analysis of the reflectance data, see
Fig.~\ref{Sr-Phononen}. We identify eight peaks of IR active
transverse-optical (TO) phonon modes, corresponding to
R$\bar{3}$c-symmetry. A theoretical analysis with
R$\bar{3}$c-symmetry has been performed by Abrashev \textit{et
al.} \cite{Abrashev99} for the phonon frequencies of LaMnO$_{3}$.
In Tab.~\ref{tabmoden} we present a comparison between the
experimental values of the IR active phonon frequencies of LSMO
\#2 at $T=250$~K and the calculated frequencies. The calculated
positions for the undoped compound can only be taken as a rough
estimate for LSMO since the lattice constants and atomic positions
are modified  by Sr-doping. Also the assumption of equal Mn-O bond
lengths in the MnO$_{6}$ octahedra\cite{Abrashev99} is not
strictly valid for the doped compounds, since the JT effect breaks
the symmetry dynamically and thereby affects the groups of bending
and stretching modes. Finally, the unequal masses of the
(La/Sr)-ions should influence the external
mode.\cite{MacManus00,Mayr03} Several resonance frequencies for
La$_{0.7}$Sr$_{0.3}$MnO$_{3}$ are known from neutron scattering
measurements, which are in good agreement with our
data.\cite{Reichardt99}

\begin{figure}[t]
\vspace{0mm} \centering
\includegraphics[width=.47\textwidth,clip,angle=0]{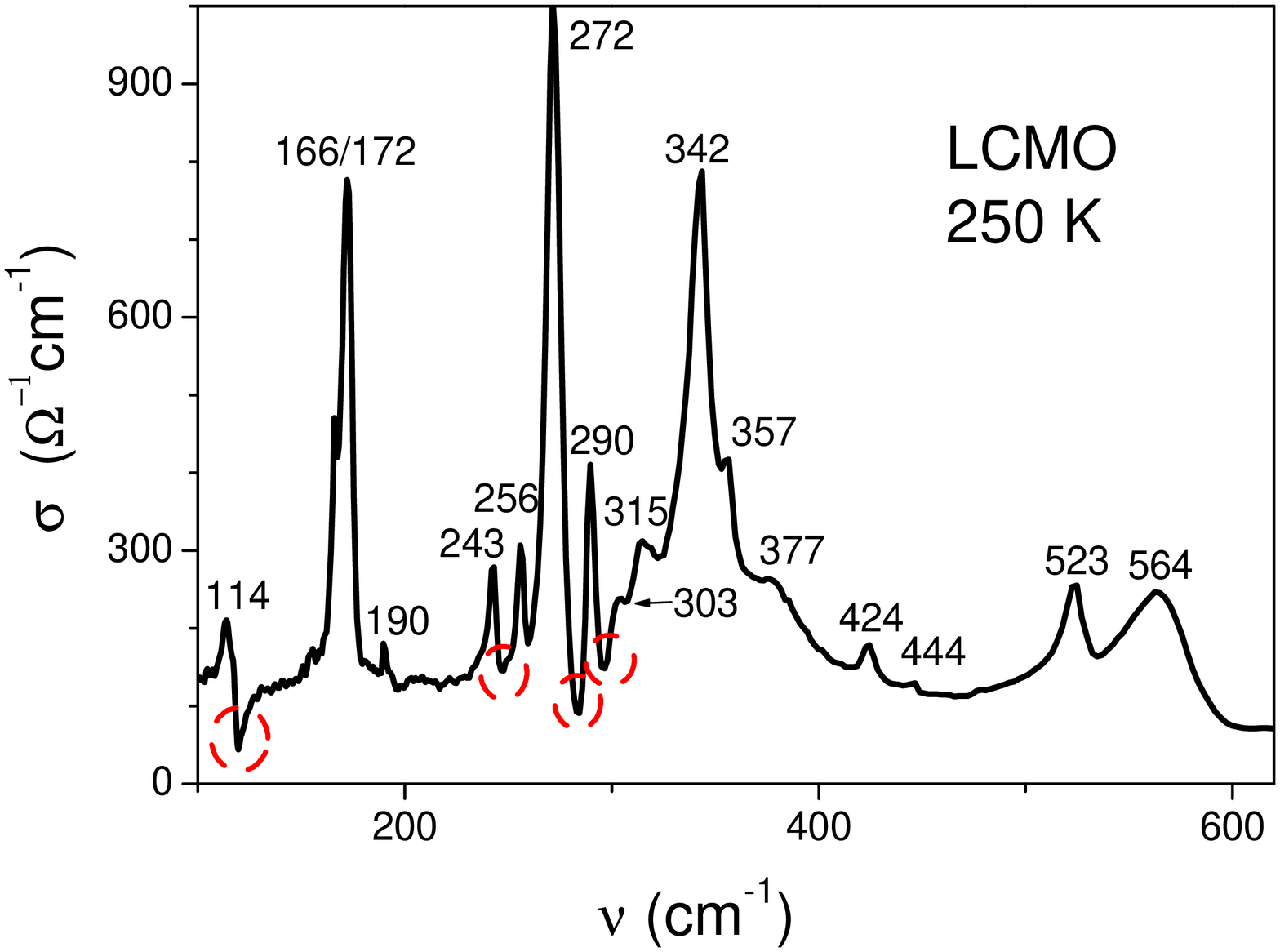}
\vspace{0mm} \caption[]{\label{Ca-Phononen}Optical conductivity of
the LCMO film at 250~K. The numbers correspond to the phonon
frequencies. The circles mark anomalies related to a strong
electron-phonon coupling (see Sec.~\ref{sec:Fano}).}
\end{figure}

In Fig.~\ref{Ca-Phononen} the optical conductivity of LCMO at
250~K is plotted jointly with labels for the phonon
eigenfrequencies. The latter result from a fit with noninteracting
harmonic oscillators. From the series of observed sharp peaks
below $\sim$600~cm$^{-1}$ we can clearly identify 17 phonon modes.
The theoretical studies for the Pnma symmetry performed by Fedorov
\textit{et al.}\cite{Fedorov99} and Smirnova\cite{Smirnova99} were
evaluated for the undoped case. Both sets of theoretical
frequencies are remarkably different from the experimental data
reported in detail by Paolone \textit{et al.}\cite{Paolone00} for
LaMnO$_{3}$.

\begin{table}[b]
\caption[]{Comparison of the phonon resonance frequencies (in
cm$^{-1}$) at 250~K and 6~K for the LCMO film with results by
Paolone \textit{et al.} \cite{Paolone00}  for pure LaMnO$_{3}$ at
300~K (P 300~K) und at 10~K (P 10~K).} \label{tabmoden-ca}
\vspace{.2cm} \centering
\begin{tabular}{cc||cccc|ccccc} \hline\hline
Osc. Nr.   && P 10~K    &&  6~K    &&  P 300~K  && 250~K\\ \hline
  1  &&  116   &&     &&  114    &&  114   &&    \\
  2  &&  119.5   &&     &&      &&     &&    \\
  3  && 162    && 166    &&      && 166    &&    \\
  4  && 172    && 175    &&  172    && 172    &&    \\
  5  && 183.5    && 187    && 181.5     &&  190   &&    \\
  6  && 201    &&     && 199     &&     &&   \\
  7  && 207    &&     &&      &&     &&    \\
  8  && 245    && 244    &&  245    && 243    &&    \\
  9  && 249    &&     &&      &&     &&    \\
  10  && 268    && 259    &&      && 256    &&    \\
  11  && 275    && 273    &&  275    && 272    &&    \\
  12  && 280    && 290    &&  280    && 290    &&    \\
  13  && 287    && 305    &&  287    && 303    &&    \\
  14  && 318    && 315    &&  315    && 315    &&    \\
  15  && 340    && 344    &&      && 342    &&    \\
  16  && 352    && 355    &&  350    && 357    &&    \\
  17  && 360    && 385    &&  362    && 377    &&    \\
  18  && 400    &&     && 400     &&     &&    \\
  19  && 426    && 426    && 430     && 424    &&    \\
  20  && 434    &&     &&      &&  444   &&    \\
  21  && 449    &&     &&  450    &&     &&    \\
  22  && 478    &&     &&  475    &&     &&    \\
  23  && 512    &&  523   &&  512    && 523    &&    \\
  24  && 561    && 589    &&  563    && 589    &&    \\
  25  && 640    &&     &&  640    &&     &&    \\
\hline\hline
\end{tabular}
\end{table}

In Tab.~\ref{tabmoden-ca} the phonon frequencies for LCMO at 6~K
and 250~K are listed and compared with published data on the
infrared spectra of undoped LaMnO$_{3}$.\cite{Paolone00} At 6~K we
can identify 15 and at 250~K 17 phonon modes. With increasing
doping concentration weak phonon modes become broader and
disappear in the background. Therefore not all 25 infrared active
modes can be resolved. Remarkably, no phonon frequency larger than
600~cm$^{-1}$ have been observed in LCMO. The deviations of the resonance
frequencies in pure LaMnO$_{3}$ from those in Ca-doped films are a
consequence of a less distorted structure but enhanced
disorder.

\subsection{Phononic  resonances across the FM-PM transition}
\label{sec:comp}
\begin{figure}[b]
\vspace{4mm}
\centering
\includegraphics[width=.48\textwidth,clip,angle=0]{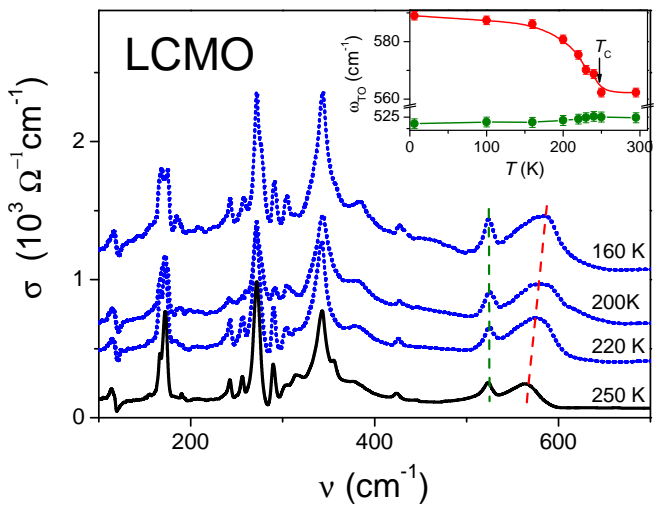}
\caption[]{\label{Ca-Min} Optical conductivity spectra of LCMO at
various temperatures; dotted lines are spectra in the FM phase,
solid lines in the PM phase. The dashed lines are to guide the
eyes for the shift of the eigenfrequencies of the stretching
modes. The higher mode changes significantly with temperature. The
inset represents the eigenfrequency shift of both stretching modes
versus temperature.}
\end{figure}

\begin{figure}[t]
\vspace{4mm}
\centering
\includegraphics[width=.48\textwidth,clip,angle=0]{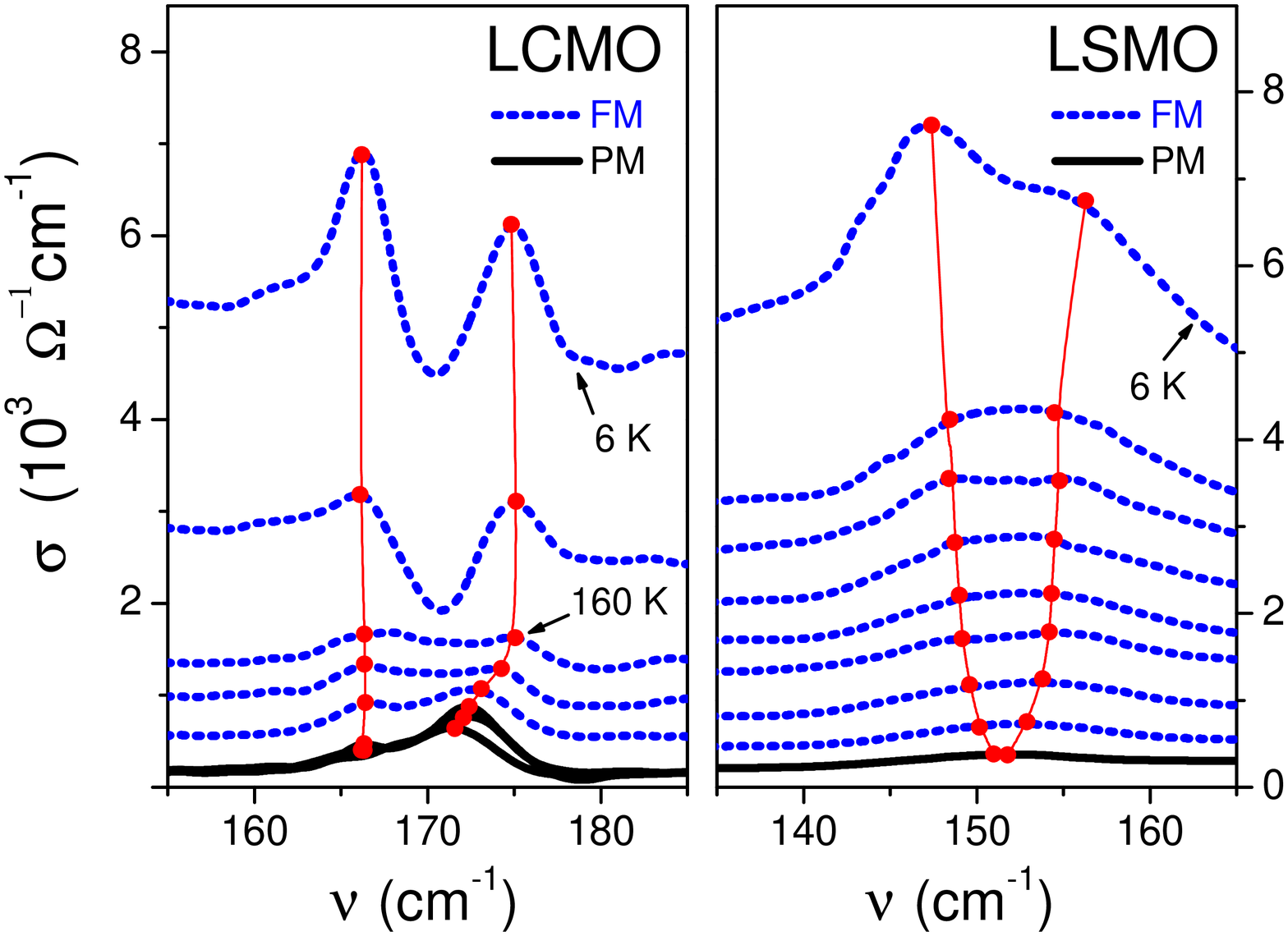}
\caption[]{Comparison of the external modes for LCMO (left) and
LSMO (right); dotted lines are spectra in the FM phase, solid
lines in the PM phase. For LCMO the shift of the high-frequency
mode near MIT is clearly visible, while for LSMO the external
modes become degenerate at about $T_{C}$. Temperatures for LCMO in
K, from top to bottom: 6, 100, 160, 200, 220, 240, 250, 295, and
for LSMO: 6, 100, 125, 160, 180, 200, 250, 295,
345.\label{external}}
\end{figure}

The spectral resolution of phononic excitations in the films is
superior to that in
bulk LCMO samples and we can resolve even both stretching modes in
the Pnma symmetry. Only the high-frequency stretching mode (see
inset of Fig.~\ref{Ca-Min}) exhibits a significant shift to lower
energy with increasing temperature. This observation is to be related
to the previously reported IR spectra of a polycrystalline LCMO sample by
Kim \textit{et al.}\cite{Kim96} The authors observed only three
broad modes of which the stretching mode  displays a notable
frequency shift around $T_C$. This shift on approaching the MIT
was attributed to a reduction of screening through mobile charge
carriers\cite{Min00} and, thereby, the authors depreciated a
magneto-elastic effect. One argument against this interpretation
is the observation that only one mode of the stretching group
would be ``sizeably less screened''\cite{comment2}---and we will see in
Fig.~\ref{external} for the external modes that again only one
mode shifts appreciably close to the MIT.

Some representative spectra of the phonon modes are seen in
Fig.~\ref{Ca-Min}. Remarkably, at low temperatures a sharply
resolved phononic structure remains which indicates that screening
effects by mobile charge carriers play only a minor role.
Consequently, we propose that the frequency shift of the
high-frequency stretching mode (see inset of Fig.~\ref{Ca-Min})
has a different origin. In this regard, the importance of
spin-lattice correlations has been pointed out by Podobedov
\textit{et al.},\cite{Podobedov98} who observed similar
temperature effects of the phonon modes through polarized Raman
spectra in an undoped sample (i.e.\ metallic screening is
absent).~\cite{comment0} Their assignment of the shift to
spin-lattice correlations is in agreement with x-ray absorption
fine-structure measurements for La$_{1-x}$Ca$_{x}$MnO$_{3}$ by
Booth \textit{et al.}\cite{Booth96} In the following we will
further elucidate the correlation between the anomalous shift and
the magnetization.

An intriguing example of a mode which exhibits the addressed
anomaly in LCMO is the external mode (see the reflectivity in the
inset of Fig.~\ref{Ca-r-temp} and the optical conductivity in
Fig.~\ref{external}, left panel). The lowering of the symmetry
from purely cubic to orthorhombic or rhombohedral implies a
diversification of the three main phonon modes (external, bending
and stretching) with the result that also the external mode
depends on the Mn-O geometry.\cite{Smirnova99}

As a consequence of the anharmonicity of the lattice an increase
of the phonon resonance frequencies with decreasing temperature is
expected.\cite{Wallis62} However, the considered phonon modes show
an irregular temperature dependence (see Fig.~\ref{external} for
the external modes, and the inset of Fig.~\ref{Ca-Min} for the
stretching modes). The shifts are controlled by the phase
transition at 245~K. In the left panel of Fig.~\ref{external} the
positions of both external modes of LCMO are displayed for various
temperatures. The lower mode exhibits only slight variations
whereas the higher mode shows a clear anomaly which sets in at
about 160~K.

\begin{figure}[t]
\centering
\includegraphics[width=.48\textwidth,clip,angle=0]{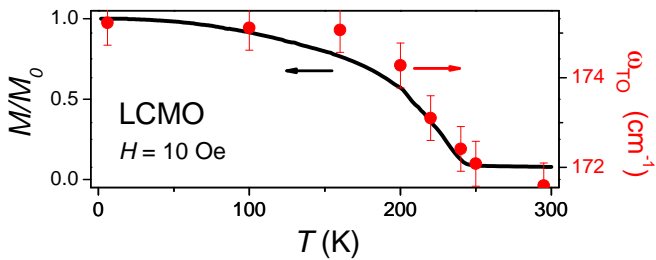}
\vspace{0mm} \caption[]{ Eigenfrequency of the high-energy
external mode at various temperatures for LCMO (right scale). The
solid line is the normalized magnetization (left scale).
\label{CaSquid}}
\end{figure}

The observation that the magnetization (solid line in
Fig.~\ref{CaSquid}) scales with the frequency shift of the higher
external mode (solid circles) below $T_C$, suggests a correlation
between spin polarization and this external mode. In comparison,
the frequency shift of the high-energy stretching mode in the
inset of Fig.~\ref{Ca-Min} is seven times larger. As expected, the
external mode is only slightly affected by the local,
correlation-induced distortion of the Mn-O\ octaedra.
We point out that a link between magnetic correlations and
phononic frequency shifts was realized
before\cite{Baltensperger70} and it was applied successfully to
explain the behavior of phonon modes in a ferromagnetic
spinel.\cite{Wakamura76}

The right panel of Fig.~\ref{external} presents the two external
modes of LSMO. While the low-energy mode shows a softening with
decreasing $T$, the higher mode displays the previously discussed
hardening. Overall, the modes approach each other, and at $T_{C}$
hardly two modes are resolvable.~\cite{Hartinger04b} We infer from
this observation that the temperature scale of the mode evolution
is in fact the magnetic temperature scale $T_C$.

Contrary to their behavior in LSMO, the two external modes in LCMO
do not become degenerate when $T_C$ is approached. However in the next section, we will
understand that electron-phonon coupling is much stronger in LCMO
with consequences for the mode structure. In fact, interference of
nearby modes is pronounced in LCMO (see Fig.~\ref{Ca-Fano}). We
will discuss this observation within a phenomenological approach
although, without a detailed microscopic analysis, it is not
feasible to fix position and weight of the considered phononic
resonances conclusively. The formation of the different phononic
resonances is strongly influenced by the details of electronic
correlations and their coupling to the phonons. Nevertheless, we
will see that for some of the modes, which strongly interact with
the electronic correlations, a considerable amount of their weight
may have its origin in a transfer of spectral weight from the
electronic degrees of freedom (cf.\ the ``bare'' mode-amplitude
$A_2$ in Tab.~\ref{doubleMode} with the resulting amplitude in
Fig.~\ref{Ca-Fano} for the lower external mode at about
170~cm$^{-1}$). It is conceivable that differences in the phononic
mode structure of LSMO and LCMO---the shape and especially the
relative weights of the phononic resonances---are, to a large
extent, controlled by electron-phonon coupling.

\section{Fano Lineshapes in LCMO}
\label{sec:Fano}

\begin{figure}[t]
\centering
\includegraphics[width=.4\textwidth,clip,angle=0]{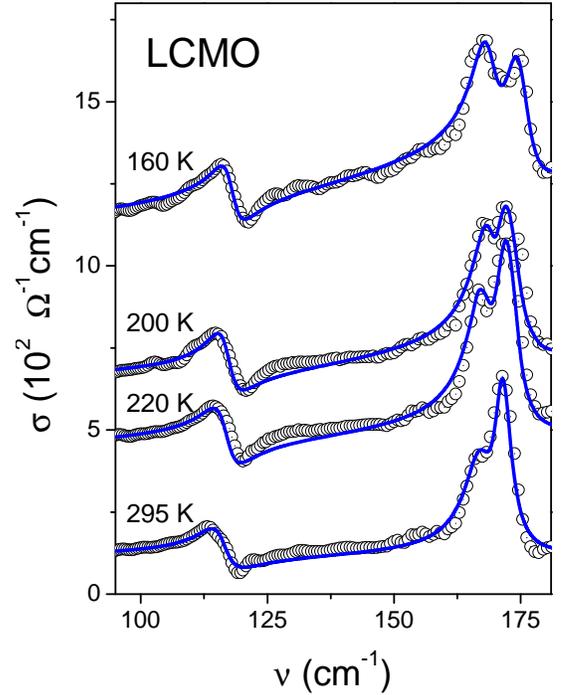}
\vspace{0mm} \caption[]{\label{Ca-Fano} Fano line shape of the
phonon mode at about 114~cm$^{-1}$  and interference of the nearly
degenerate external modes at about 170~cm$^{-1}$ in LCMO. Fit with
Eqs.~(\ref{Amplitude_SP})--(\ref{GD}) (line) to experimental data
(open circles) at various temperatures.}
\end{figure}

The lineshape of the phonon spectrum evidences the strength of the
electron-lattice correlations  as the coupling  of lattice
vibrations to excitations of the electronic system may lead either
to destructive or constructive interference. The Lorentzian
resonant shape of the phonon mode is thereby deformed into a
resonance anti-resonance (peak-dip) structure if, for example, the
phase of the electronic continuum excitations is nearly constant
in the frequency range where the phononic excitation experiences a
phase shift by $\pi$. The detailed form of the resonant structure
then depends on the coupling strengths and the amplitudes of the
considered excitations. A beautifully balanced peak-dip structure
is observed for the lowest external mode at about 114~cm$^{-1}$ in
LCMO (see Fig.~\ref{Ca-Fano}). Such an asymmetric Fano line
shape\cite{Fano61,Hackl99}  indicates a sizeable interaction
between lattice vibrations and the electronic continuum. The
asymmetry is also seen for other phonon modes (marked by open
circles in Fig.~\ref{Ca-Phononen}). We observe these Fano line
shapes in LCMO, not in LSMO, where the electron-lattice coupling
has to be considerably weaker.

Since a microscopic calculation of the resonance shape in these
systems with charge, spin, orbital and lattice degrees of freedom
is not yet feasible we discuss the asymmetry and
coupling-dependent position and weight of the external modes
within a phenomenological approach. It accounts for the phase
relation between various oscillator modes which effectively model
phonons and electronic continua. The coupling of these modes not
only leads to Fano-like interference effects but also to
mode splitting and transfer of spectral weight. The parameters
which enter the model (amplitudes, eigenfrequencies, relaxation
rates and couplings) will be denoted as ``bare parameters'' and
they will be determined from an optimal fit to the considered
spectra. However it should be apprehended that they cannot be the
bare parameters of a microscopic theory but rather contain
residual interactions since the excitation spectrum cannot be
fully decomposed into independent modes which couple in a scalar,
linear way as assumed in the phenomenological modelling.

We analyze the present LCMO data by applying the phenomenological
approach of Burlakov \textit{et al.},\cite{Burlakov92,Zibold92}
taking into account several adjacent phonons which interact with
electronic continua. The advantage of this approach is the
explicit specification of and control over the electronic
continua. In reference to our results in the mid-infrared, we take
a small polaron (SP) resonance peaked between 1990~cm$^{-1}$ at
160~K and 3300~cm$^{-1}$ at room temperature\cite{Hartinger04a}
Furthermore, we introduce an electronic oscillator in the
far-infrared range, which has to be related to an incoherent
contribution of orbital excitations\cite{Horsch99, Shiba97,
Ishihara97} in  a unified treatment of all electronic and phononic
degrees of freedom. As this cannot be the scope of the paper we
identify these excitations phenomenologically with the spectral
function of Eq.~(\ref{Ge}) (see below). This ``electronic
oscillator'' contribution is peaked around 300~cm$^{-1}$ in our
optical measurements and it will be referred to as ``incoherent
contribution''. We also include a Drude term which accounts for
the small frequency side of the FIR spectrum but has only marginal
influence on the phonon modes.

For the evaluation of the optical conductivity $\sigma(\nu)$ we
introduce a resonant interaction of the phonons with the electronic
excitations:\cite{Burlakov92,Zibold92}

\begin{equation}\label{Amplitude_SP}
\textrm{Re} \, \sigma(\nu)= \varepsilon_{0} 2\pi c
\,\textrm{Im}[\hat{A}\hat{G}(\nu)\hat{A}]
\end{equation}

\noindent where
\begin{equation}\label{AAA}
 \hat{A}=(A_{1},...,A_{n},A_{sp},A_{e},A_{d})
\end{equation}
\noindent is a vector of $n$ matrix elements ($A_{j}$) for optical
dipole transitions of phonon modes, for the small polaron
($A_{sp}$), for an incoherent ($A_{e}$) and for a Drude
contribution ($A_{d}$). $\hat{G}$ is the Green function matrix of
the electron-phonon system. Thereby $A_{sp}$ is calculated by
applying the $f$-sum rule to the SP component
\begin{equation}\label{ASP}
A_{sp}= \sqrt{\frac{1}{\varepsilon_{0} 2 \pi^{^2}
c}\int^{\infty}_0 d\nu \, {\rm Re}\,\sigma_{sp}(\nu)}
\end{equation}

\noindent where $\varepsilon_{0}$ is the vacuum permittivity and
$c$ is the speed of light. The SP optical conductivity is:\cite{Puchkov95,Yoon98}
\begin{equation}
\label{eq:polaron} {\rm Re}\,\sigma_{sp}(\nu,T)= \sigma_0(T) \frac{\sinh
(4 E_{b} \nu / \Delta^2)}{4 E_{b} \nu / \Delta^2}
\, e^{-\nu^{2}/\Delta^{2}}  \quad
\end{equation}
Here $\sigma_0(T)$ is the dc-conductivity, $E_{b}$ is the SP
binding energy, $\Delta\equiv 2 \sqrt{2 E_{b} E_{\rm vib}}$, and
$E_{\rm vib}$ is the characteristic vibrational energy which is
the thermal energy $T$ in the high-temperature regime and
$\nu_{\rm ph}/2$ at low temperatures ($k_B T < \hbar 2\pi
c\nu_{\rm ph}$, $\nu_{\rm ph}$ is a phonon frequency). The
dimension of $\nu$ and all other energy scales is cm$^{-1}$.

We assume that not only the phonon modes approximately decouple
but also the electronic modes, which generate the SP resonance,
the Drude contribution and the resonance in the FIR range. The
inverse Green function matrix is in this phenomenological scheme
\begin{equation}\label{GFMatrix}
\hat{G}^{-1}=  \begin{pmatrix}
    G^{-1}_{1} & 0 & 0 &\ldots &g_{1} & k_{1}& 0\\
    0 & G^{-1}_{2} & 0 &\ldots& g_{2} & k_{2}&0 \\
    0 & 0 & G^{-1}_{3} &\ldots& g_{3} & k_{3}& 0 \\
    \vdots & \vdots & \vdots &\ddots& \vdots & \vdots  &\vdots\\
    g_{1} & g_{2} & g_{3} &\ldots& G^{-1}_{sp} & 0&0\\
    k_{1} & k_{2} & k_{3} &\ldots& 0 & G^{-1}_{e}&0 \\
    0 & 0 & 0 &\ldots& 0 &0 & G^{-1}_{d} \
  \end{pmatrix}
\end{equation}
\noindent with the respective Green functions of the phonon modes,
the small polaron, incoherent and the Drude background:
\begin{equation}\label{GPH}
G_{j}(\nu)=\frac{\nu}{\nu_{j}^2-\nu^{2}-i\nu\gamma_{j}}\,
\end{equation}
\begin{equation}\label{GSP}
G_{sp}(\nu)=\frac{i
\,\textrm{Re}\,\sigma_{sp}(\nu)-\textrm{Im}\,\sigma_{sp}(\nu)}{\varepsilon_{0}
2\pi c \,A_{sp}^2}\,
\end{equation}
\begin{equation}\label{Ge}
G_{e}(\nu)=\frac{\nu}{\nu_{e}^2-\nu^{2}-i\nu\gamma_{e}}\,
\end{equation}
\begin{equation}\label{GD}
G_{d}(\nu)=-\frac{1}{\nu+i\gamma_{d}}\,
\end{equation}
\noindent $\gamma_{j}$ and $\nu_{j}$ are the width and the
frequency of the $j$th phonon mode, $\gamma_{e}$ and $\nu_{e}$ of
the electronic oscillator, and $\gamma_{d}$ parameterizes the
Drude contribution. The imaginary part of $\sigma_{sp}$ is
obtained from a Kramers-Kronig relation. The strengths of the
polaron-phonon ($g_{j}$) and electron-phonon ($k_{j}$) coupling
parameters determine the asymmetric line shape while the sign
controls at which side of the phonon mode the dip will appear. For
$g_{j}=k_{j}=0$ a Lorentzian shape is recovered for the
jth~phonon.\cite{comment3} In relation~(\ref{GFMatrix}) we set the
coupling of the Drude-phonon coupling to zero as its effect on the
resonant lineshapes is small for the considered temperature range
below the MIT and we thereby avoid a larger number of fitting
parameters. Moreover the Drude contribution is absent above the
MIT but the lineshapes are still similar to those below the MIT,
an observation which excludes a priori a dominant role of the
Drude term.

The fit procedure consists of the following three steps: First,
the polaron parameters are fixed through the fit of the
mid-infrared polaron resonance.\cite{Hartinger04a} Then the
background in the FIR range is parameterized through $G_e$ and
$G_d$. Finally, the frequencies and widths of the phonon
resonances are determined, the coupling constants are extracted
from fitting their respective shape, and the frequencies and
widths are readjusted in order to gain the optimal fit.

In Fig.~\ref{Ca-Fano} we show experimental and theoretical data
for the low-frequency phonon group measured at different
temperatures. We now analyze in detail the most distinctive
asymmetric mode in LCMO at about 114~cm$^{-1}$ and include the two
adjacent phonon modes for the interpretation. This group is well
separated from other phonon excitations, which allows us to
neglect the coupling to the remaining modes. In
Tab.~\ref{Tab-Ca-Fano} the interaction parameters  on the basis of
our approach  are presented for the 114~cm$^{-1}$ phonon for
different temperatures~$T$.

\begin{table}[t]
\caption[]{Parameters of the Fano-like fit for the 114~cm$^{-1}$
phonon of LCMO  at different $T$ using the approach of
Eqs.~(\ref{Amplitude_SP})--(\ref{GD}).} \label{Tab-Ca-Fano}
\vspace{.2cm} \centering
\begin{tabular}{lc|ccccccccccccc}
 \hline \hline
T (K) &&& 160 &&& 200 &&& 220 &&& 295 \\
 \hline
 $A_{1}$ (cm$^{-1}$)&&& 150 &&& 148 &&& 143 &&& 140 \\
 $\gamma_{1}$ (cm$^{-1}$)&&& 4.1 &&& 4.1 &&& 4.1 &&& 3.9 \\
 $\nu_{1}$ (cm$^{-1}$)&&& 118.0 &&& 117.2&&& 116.9 &&& 116.4 \\
 $g_{1}$ (cm$^{-1}$)&&& -42  &&& -70 &&& -115 &&&-280\\
 $k_{1}$ (cm$^{-1}$)&&& -3 &&& -8 &&& -9 &&&-16\\\hline
 $A_{sp}$ (cm$^{-1}$)&&& 16502&&& 15251 &&& 15958 &&& 15495 \\
 $\sigma_0$ ($\Omega^{-1}$cm$^{-1}$)&&& 890&&& 460 &&& 350 &&& 108 \\
 $E_b$ (cm$^{-1}$)&&& 1990&&& 2000 &&& 2520 &&& 3300 \\
 $\nu_{\rm ph}$ (cm$^{-1}$)&&& 350&&& 261 &&& 310 &&& 300\\\hline
 $A_{d}$ (cm$^{-1}$) &&& 1350 &&& 1350&&& 1350 &&& -- \\
 $\gamma_{d}$ (cm$^{-1}$)&&& 140 &&& 150 &&& 350 &&& -- \\\hline
 $A_{e}$ (cm$^{-1}$) &&& 3500 &&& 3370&&& 2800 &&& 1600 \\
 $\gamma_{e}$ (cm$^{-1}$)&&& 370 &&& 340 &&& 280 &&& 250 \\
 $\nu_{e}$ (cm$^{-1}$)&&& 276 &&& 320 &&& 340 &&& 345 \\
 \hline \hline
\end{tabular}
\end{table}

\begin{table}[b]
\caption[]{Parameters of the $T$-dependent strong-coupling fit for
the two external phonon modes of LCMO at about 170~cm$^{-1}$, using the
approach of Eqs.~(\ref{Amplitude_SP})--(\ref{GD}). Note that the frequencies in the
table are the bare frequencies which differ from the
interaction-dependent frequencies of the resonances in the optical
conductivity, cf.~Tab.~\ref{tabmoden-ca}.} \label{doubleMode}
\vspace{.2cm} \centering
\begin{tabular}{lc|ccccccccccccc}
 \hline \hline
T (K) &&& 160 &&& 200 &&& 220 &&& 295 \\
 \hline
 $A_{2}$ (cm$^{-1}$)&&& 220 &&& 120 &&& 52 &&& 10 \\
 $\gamma_{2}$ (cm$^{-1}$) &&& 5.0 &&& 4.3 &&& 3.6 &&& 3.2 \\
 $\nu_{2}$ (cm$^{-1}$)&&& 169 &&& 169.5 &&& 169.8 &&& 171.1 \\
 $g_{2}$ (cm$^{-1}$)&&& -7 &&& -3 &&& -2 &&& 0 \\
 $k_{2}$ (cm$^{-1}$)&&& -16 &&& -28 &&& -47 &&& -65 \\ \hline
 $A_{3}$  (cm$^{-1}$) &&& 290 &&& 318 &&& 370 &&& 380 \\
 $\gamma_{3}$ (cm$^{-1}$)&&& 5.0 &&& 4.7 &&& 4.5 &&& 4.2 \\
 $\nu_{3}$ (cm$^{-1}$)&&& 175.0 &&& 172.8 &&& 172.5 &&& 171.5 \\
 $g_{3}$ (cm$^{-1}$)&&& -9 &&& -4 &&& -2 &&& 0 \\
 $k_{3}$ (cm$^{-1}$)&&& -12 &&& -14 &&& -22 &&& -28 \\
 \hline \hline
\end{tabular}
\end{table}

All line shapes of the first phonon mode in Fig.~\ref{Ca-Fano}
exhibit a dip on the high-energy side, a consequence of the
negative sign of the coupling parameters. The coupling constant is
proportional to the effective mass. In manganites the carriers
have hole character.\cite{Ju97,Saitho95} The associated inverse
sign of the effective mass may be responsible for a negative value
of $g_{j}$ and $k_{j}$. It turns out that neither a  coupling to
the Drude nor to the incoherent contribution can induce the nearly
perfect antisymmetric shape of the 114~cm$^{-1}$ peak-dip
structure. Only a sufficiently strong coupling to the polaronic
excitation supports this antisymmetric Fano shape. The increase of
coupling strength $g_1$ with temperature is significant. It
results from the fact that the line shape keeps its antisymmetric
form up to the highest temperature (295~K) while the electronic
background is reduced by a factor 8  from 160~K to 295~K. Since
the line shape of the phonon is controlled by a product of the
coupling constant and the strength of the polaronic background
($\sigma_0$), the coupling constant has to increase
correspondingly to keep the observed shape.

\begin{figure}[t]
\centering
\includegraphics[width=.4\textwidth,clip,angle=0]{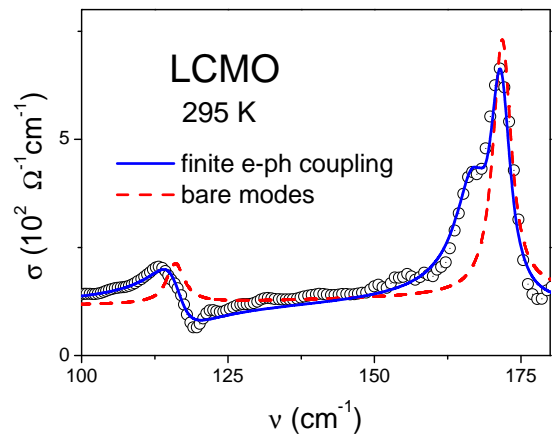}
\vspace{0mm} \caption[]{\label{Ca_Fano_vgl_T295} Fits of the low
frequency phonon group with and without electron-phonon (e-ph) coupling
for data at room temperature.}
\end{figure}

The line shapes of the two modes at about 170~cm$^{-1}$ do not
deviate strongly from Lorentzians which is consistent with rather
small couplings $g_2$ and $g_3$ (cf.~Tab.~\ref{doubleMode}).
However, in order to reproduce the ``hump'' in between the phonon
at 114~cm$^{-1}$ and the nearly degenerate external modes at about
170~cm$^{-1}$ one needs a sufficiently strong electron-phonon
coupling $k_{2}$ to the electronic resonance at about
300~cm$^{-1}$. These electron-phonon couplings, $k_{2}$ and
$k_{3}$, affect the position and the spectral weight of the
resonances at $\nu_{2}$ and $\nu_{3}$ considerably. The three
coupled modes ($\nu_{2}$, $\nu_{3}$, $\nu_{e}$) constitute a
``triad'' with a bonding (lowest), non-bonding (intermediate) and
anti-bonding state. The bonding and anti-bonding modes always
experience level repulsion so that the bare frequencies $\nu_{2}$
and $\nu_{3}$ can be both at 171~cm$^{-1}$ (degenerate modes,
cf.~Tab.~\ref{doubleMode}) but the two resonances are still well
separated at 295~K (at 167~cm$^{-1}$ and at 171~cm$^{-1}$, cf.~
Fig.~\ref{Ca-Fano} and Tab.~\ref{tabmoden-ca}).~\cite{comment4}
The level repulsion is accompanied by a transfer of spectral
weight: according to Tab.~\ref{doubleMode} and Fig.~\ref{Ca-Fano},
the lowest mode of the triad takes nearly all its weight from the
electronic mode at 295~K. This observation may explain why the
two-mode structure at 170~cm$^{-1}$ in LCMO is still intense at or
above the MIT whereas in LSMO the mode at about 170~cm$^{-1}$ is
weak (cf.\ Fig.~\ref{external}, right panel). In
Fig.~\ref{Ca_Fano_vgl_T295} we display the ``bare modes'', with
parameters of Tabs.~\ref{Tab-Ca-Fano} and \ref{doubleMode},
jointly with the interaction-dominated mode spectrum in order to
summarize visually the effects of the strong electron-coupling on
the external mode group in LCMO.

In contrast to LCMO, the phonon spectra for LSMO exhibit no
asymmetric shape and can be described by Lorentzian oscillators.
In the latter compound the electron-phonon coupling is smaller, as
is confirmed by shape, position and temperature dependence of the
polaronic mid-infrared (MIR) resonance.\cite{Hartinger04a}

\section{Conclusions}
\label{sec:conclusion}

Manganite thin films provide an excellent opportunity to study the
phononic excitations of LSMO and LCMO in their metallic phases.
The films allow for a well-resolved, distinct phonon spectrum in
infrared spectroscopy, even for temperatures well below the MIT.
In LSMO we identified each of the eight infrared-active phonons
expected for the R$\bar{3}$c symmetry and in LCMO we were able to
resolve 18 out of 25 phonons expected for the Pnma-symmetry.

The phonon spectra reveal a number of anomalies which result from
a finite electron-lattice coupling and which confirm the
importance of cooperative effects of microscopic degrees of
freedom when approaching the ferromagnetic-to-paramagnetic and the
metal-insulator transition.

First and most apparent, we observe that the FIR spectra change
substantially up to the MIT (see Figs.~\ref{Ca-r-temp} and
\ref{Sr-r-temp}) but then, above the MIT, temperature dependent
modifications are negligible. This overall $T$-dependence of the
spectra up to the MIT is  related to a strongly reduced screening
of the phononic excitations, due to localization or diffusive
motion of electronic charges. However also a coupling of lattice
and spin degrees of freedom is evidenced by the phononic
excitation spectrum: temperature dependent shifts of several
phonon modes, which vary on the same temperature scale as the
magnetization, confirm the correlation of these degrees of freedom
(see Figs.~\ref{Ca-Min} -- \ref{CaSquid}).

In LCMO the pronounced asymmetric line shapes, observed for
several phonons (see Figs.~\ref{Ca-Phononen} and \ref{Ca-Fano}),
are a manifestation of the strong electron-phonon coupling. We
analyzed the Fano lineshapes of the external group within a
phenomenological approach.~\cite{Burlakov92,Zibold92} It assigns
independent modes to the various spectral contributions of the
electronic excitations and couples them linearly to the phononic
oscillators. In FIR and MIR spectra of LCMO, three electronic
continua are typically identified: a ``continuum of incoherent
electronic excitations'' (cf.~Refs.~\onlinecite{Horsch99, Shiba97,
Ishihara97}), observed as a background to the phonons in the FIR
range, polaronic excitations in the MIR and possibly a rather
insignificant Drude contribution on the low-frequency side of the
FIR spectrum.

The lineshape of the lowest-frequency external mode, which is
remarkably antisymmetric at all temperatures, can be reproduced
only with a strong coupling to the polaronic excitations. In
contrast, the FIR continuum, which is parameterized by a strongly
damped oscillator spectral function, is responsible for frequency
shifts (level repulsion) and transfer of spectral weight from the
electronic excitations into distinct phononic modes (see
Fig.~\ref{Ca_Fano_vgl_T295}). In the presented scheme, we find
that the mode splitting of the 170~cm$^{-1}$ resonance and its
considerable spectral weight at high temperature (above the MIT)
is induced by the electronic continuum. This observation may
explain the discrepancy with LSMO where these two external modes
are degenerate at high temperature since LSMO is characterized by
a weaker electron-phonon coupling where Fano-like lineshapes are
entirely missing. A microscopic evaluation is mandatory in order
to assign unambiguously these anomalous characteristics of the
phononic spectra to the interference with specific electronic
excitations. The preliminary identification within the
phenomenological approach has provided ample motivation for such
an effort.

\begin{acknowledgments}
The  research was supported by  BMBF
(13N6917, 13N6918A) and by DFG through the Sonderforschungsbereich
SFB 484 (Augsburg).
\end{acknowledgments}


\begin{thebibliography}{1}
\bibitem{comment2} Moreover, a full
evaluation of screening would have to include the diffusive nature
of the charge transport which results in a retarded
screening dynamics with a less prominent signature.
\bibitem{comment0} Near the magnetic transition an irregular shift at
    609~cm$^{-1}$ was attributed to spin-lattice interaction due to
    the observed variations of the lattice parameters and sublattice
    magnetization.
\bibitem{comment3} In the original Fano model a parameter $q$ determines the degree
   of the line asymmetry of the peak, which is inversely related to
   the strength of interaction.\cite{Damascelli97,Lupi98}
\bibitem{comment4} Although the scheme with a pure triad of bonding,
    non-bonding and anti-bonding modes is oversimplified, it visualizes sufficiently
    the expected modifications of mode positions and spectral weight
    with increasing coupling. Certainly, the higher-energy external mode
    at about 172~cm$^{-1}$ has not a pure non-bonding character and
    it also experiences a small level shift and modifications of its
    spectral shape through its interaction with the other modes.
\end{thebibliography}
\end{document}